\documentclass[12pt]{article}
\usepackage{pic02}
\usepackage{hyperref}
\usepackage{url}
\usepackage{graphicx}
\input {babarsym.tex}
\def\Brec{\ensuremath{B_{rec}}\xspace}
\def\Btag{\ensuremath{B_{tag}}\xspace}

\begin{document}

\def\leptontag{{\tt Lepton}}
\def\kaonitag{{\tt Kaon\,I}}
\def\kaoniitag{{\tt Kaon\,II}}
\def\othertag{{\tt Inclusive}}
% Some other macros used in the sample text

\def\st{\scriptstyle}
\def\sst{\scriptscriptstyle}
\def\mco{\multicolumn}
\def\epp{\epsilon^{\prime}}
\def\vep{\varepsilon}
\def\ra{\rightarrow}
\def\ppg{\pi^+\pi^-\gamma}
\def\vp{{\bf p}}
\def\ko{K^0}
\def\kb{\bar{K^0}}
\def\al{\alpha}
\def\ab{\bar{\alpha}}
\def\be{\begin{equation}}
\def\ee{\end{equation}}
\def\bea{\begin{eqnarray}}
\def\eea{\end{eqnarray}}
\def\CPbar{\hbox{{\rm CP}\hskip-1.80em{/}}}
\def\stwofone{\ensuremath{\sin\! 2 \phi_1}}
\def\vubvcb{\ensuremath{|{V_{ub}}/{V_{cb}}|}}
\def\deltams{\ensuremath{{\rm \Delta}m_s}\xspace}
\def\ddstarh{\ensuremath{D^{(*)-}h}\xspace}
\def\lp{\ensuremath{l^{+}}\xspace}
\def\rhop{\ensuremath{\rho^{+}}\xspace}
\def\aonep{\ensuremath{a_1^{+}}\xspace}

\def\invps   {\ensuremath{\mbox{\,ps}^{-1}}\xspace}
\def\prl{Phys. Rev. Lett. } 
\def\prd{Phys. Rev. D }
%%%%%%%%%%%%%%%%%%%%%%%%%%%%%%%%%%%%%%%%%%%%%%%%%%%%%%%%%%%%%%%%%%%%%%%%
%%% Editorial Commentary
%%%%%%%%%%%%%%%%%%%%%%%%%%%%%%%%%%%%%%%%%%%%%%%%%%%%%%%%%%%%%%%%%%%%%%%%

\noindent%

\title{\bf MEASUREMENT OF B MIXING FREQUENCY AND CP VIOLATION PARAMETER $\stwob$ ($\stwofone$) AT B FACTORY EXPERIMENTS}
\author{
Yibin Pan       \\
{\em University of Wisconsin-Madison}}
\maketitle

%
% photograph of author
%  This is where we will insert a photograph. To see what it would look like,
%  uncomment the following lines.
%
%\begin{figure}[h]
%\begin{center}
%
% include photograph for proceeding version
%
%\includegraphics[height=4.5cm]{einstein.eps}
%
% insert a fixed vertical spacing instead for the ArXiv preprint
%
\vspace{4.5cm}
%
%\end{center}
%\end{figure}

\baselineskip=14.5pt
\begin{abstract}
Recent results on B mixing and CP violation from the B-factory experiments, $\babar$ at PEP-II and Belle at KEK-B, are summarized.  A discussion of CP violation is then presented which concentrates on the CP parameter $\stwob$ (also known as $\stwofone$). The most recent measurements of this parameter from B-factory data yield 
\begin{eqnarray*}
\stwob &=& 0.741 \pm 0.067\ \stat \pm 0.033\ \syst \, \mathrm{(BaBar)}  \\
\stwofone &=& 0.719 \pm 0.072\ \stat \pm 0.035\ \syst \, \mathrm{(Belle)}. 
\end{eqnarray*}
These two B-factory results contribute to the current world average of 
\begin{eqnarray*}
\stwob &=& 0.735 \pm 0.055
\end{eqnarray*}

\end{abstract}
\newpage

\baselineskip=17pt

\section{Introduction}
In the Standard Model, CP violation is made possible by an irreducible complex
 phase in the three-generation Cabibbo-Kobayashi-Maskawa(CKM) quark-mixing 
matrix~\cite{bib:CKM}. CP violation is expected if this complex phase is non-zero. 
The unitarity of the CKM matrix results in six triangles of equal area in the 
complex plane (Unitarity Triangles). A non-zero area implies the existence of 
a CP-violating phase -- in other words, {\em if the inner angles of these 
Unitary Triangles are found to be non-zero, CP-violation is observed}.
For the neutral B-meson system, the angles in question are $\alpha$, $\beta$ 
and $\gamma$~\footnote{BaBar uses  notation \{$\alpha, \beta, \gamma$\} and Belle uses notation \{$\phi_2, \phi_1, \phi_3$\} to address same three CP angles. In this paper, $\babar$ notation is used except for description of Belle-specific results.} and are defined by the condition 
\begin{eqnarray}
V_{\rm ud}^{}V_{\rm ub}^*+ V_{\rm cd}^{}V_{\rm cb}^* + V_{\rm td}^{}V_{\rm tb}^* &=& 0.
\end{eqnarray}
In particular, the measurement of the angle $\beta$, defined as 
\begin{eqnarray}
\beta \equiv \arg \left[\, -V_{\rm cd}^{}V_{\rm cb}^* / V_{\rm td}^{}V_{\rm tb}^*\, \right],
\end{eqnarray}
is the main focus of this paper. 

Observation of CP violation in neutral $B$ decays\footnote{Charge-conjugation is implied
throughout this document.} through the measurement 
of $\stwob$ was reported in summer 2001 by both the $\babar$~\cite{bib:babar-stwob-2001} and Belle~\cite{bib:belle-stwob-2001} 
collaborations employing data luminosities of 29.7 and 29.1 $\invfb$, respectively. 
By summer 2002, each experiment's total accumulated luminosity had almost tripled.
The more precise $\stwob$ results presented in this paper are based on the full 
samples obtained for summer 2002: 81 $\invfb$ (88 million  $\BB$ decays) for $\babar$ 
and 78 $\invfb$ (85 million $\BB$ decays) for Belle.

%%%%%%%%%%%%%%%%%%%%%%%%%%%%%%%%%%%%%%%%%%%%%%%%%%%%%%%%%%%%%%%%%%%%%%%%
%%%%%%%%%%%%%%%%%%%%%%%%%%%%%%%%%%%%%%%%%%%%%%%%%%%%%%%%%%%%%%%%%%%%%%%%
%%%%%%%%%%%%%%%%%%%%%%%%%%%%%%%%%%%%%%%%%%%%%%%%%%%%%%%%%%%%%%%%%%%%%%%%

\section{B-Factory Experiments: $\babar$ and Belle}
The $\babar$ and Belle B-factory experiments share similar design concepts. 
Both experiments center around a high luminosity asymmetric electron-positron 
collider operating at the $\Y4S$ resonance. Each experiment also depends on a high-precision 
detector designed specifically for high-rate, Lorentz-boosted $\BB$ production. 
The PEP-II collider ($\babar$) counter-circulates an electron beam of 9.0 GeV against
a positron beam of 3.1 GeV, resulting in a center-of-mass boost of $\beta\gamma$ = 0.55.
The KEKB collider (Belle) collides electrons and positrons 
at 8.0 GeV and 3.5 GeV, respectively, resulting in a boost of  $\beta\gamma$ = 0.425.
B-mesons produced at each collider are boosted along the $\en$ direction 
because of the asymmetric energies,
allowing for the measurement of the decay-time difference of the two Bs.

The primary sub-detectors of $\babar$ include a drift chamber (DCH) and a silicon vertex 
tracker (SVT), both operating  inside a 1.5 T magnetic  field provided by a super-conducting solenoid.
Surrounding the tracking volume is a detector of internally reflected \v{C}erenkov radiation (DIRC), 
an electromagnetic calorimeter (EMC) and an instrumented flux return (IFR).

The Belle apparatus consists of a silicon vertex detector (SVD),
a central drift chamber (CDC), an array of
aerogel threshold \v{C}erenkov counters (ACC),
time-of-flight scintillation counters (TOF), and an electromagnetic calorimeter
(ECL)  located inside
a super-conducting solenoid coil that provides a 1.5~T
magnetic field.  An iron flux-return located outside of
the coil is instrumented to detect $K_L^0$ mesons and to identify
muons (KLM).

For the CP and mixing measurements, important detector capabilities include 
tracking, vertexing and particle identification (PID). Details 
on how each experiment address these requirements can be found in 
Ref.~\cite{bib:babar} and Ref.~\cite{bib:belle}.

%%%%%%%%%%%%%%%%%%%%%%%%%%%%%%%%%%%%%%%%%%%%%%%%%%%%%%%%%%%%%%%%%%%%%%%%
%%%%%%%%%%%%%%%%%%%%%%%%%%%%%%%%%%%%%%%%%%%%%%%%%%%%%%%%%%%%%%%%%%%%%%%%
%%%%%%%%%%%%%%%%%%%%%%%%%%%%%%%%%%%%%%%%%%%%%%%%%%%%%%%%%%%%%%%%%%%%%%%%

\section{Overview Of Measurement Technique}

\subsection{Production of $\BzBzb$}
As a consequence of Bose-Einstein statistics, the $\BzBzb$ pairs produced from 
the $\Y4S$ decays remain 
in a coherent P-wave state until one of the two B-mesons decays. At the moment 
of the first decay ($t=0$), the two B-mesons are in opposite flavor 
states -- knowing the flavor state of one B implies knowledge of the 
other.  Between the first decay and its own decay, the second B-meson's flavor 
evolves according to a time-dependent oscillatory pattern.  
If the flavor of one B is known when it decays then the flavor state of the 
other B at its decay point is solely determined by the proper time between
decays ($\deltat$) and mixing frequency $\deltamd$.  
   
\subsection{Exclusive Reconstruction of One B-Meson}
Given the large number of B-mesons produced at B-Factories, it is conceivable 
to exclusively reconstruct one of the two B-mesons into a known decay mode. 
Excluding from consideration the decay products of this reconstructed B ($\Brec$), 
the remaining particles in the event then presumably belong to the ``other B''. 
Often, the flavor of this ``other B'' can be determined through an inclusive flavor tagging 
method (B Flavor-Tagging). For this reason, this inclusively ``reconstructed'' second B-meson
is commonly referred as the ($\Btag$).  The $\Brec$ and $\Btag$ can be individually vertexed, 
and the distance between the two vertices used to determine the proper-time difference $\deltat (\equiv t_{rec}-t_{tag})$.

The choice of exclusive decay modes is determined according to  physics objectives. 
For a B-mixing measurement, the $\Brec$ has to decay into one of the (self-tagged) 
flavor eigenstates. The mixing frequency $\deltamd$ is determined by comparing the 
flavor of the $\Brec$ and $\Btag$ (both known) in a time-dependent way.  For 
CP measurements, $\Brec$'s are required to be reconstructed in a
CP eigenstate, such as $\jpsi \KS$, etc.

\subsection{$B$ Decay Time Interval}
In asymmetric $\BB$ production, as at $\babar$ and Belle,
the large boost causes the $B$ mesons to fly preferentially 
along the beam direction (conventionally the z-axis).
Accordingly, the time interval $\deltat$ between 
the two $B$ decays is calculated, to a good approximation,  as 
\begin{eqnarray}
\deltat = \deltaz/(\beta\gamma c), 
\end{eqnarray}
where $\deltaz$ is the distance between the decay vertices of $\Brec$ and 
$\Btag$ along the z-axis.  The $\Brec$ vertex 
is determined by using the charged tracks from its exclusive decay products;   
intermediate vertices, such as those from $\KS$ decay, are also reconstructed. 
The $\Btag$ vertex is obtained by an 
inclusive fit on charged tracks which do not belong to the exclusive $\Brec$. 
Constraints from the beam spot locations and $\Brec$ momentum are applied when fitting for $\Btag$.  

The $\deltat$ resolution is affected by the detector resolution for both the $\Brec$ and $\Btag$ vertices, by  a shift on the $\Btag$ vertex due to 
secondary charmed decays, and by kinematic smearing due to the fact that the $B$ flight is not exactly in the z-direction. Accordingly, an empirical resolution function is used to model these effects. In both experiments, the parameters in 
the resolution functions are determined in data from fits to the neutral and 
charged $B$ meson lifetime. An average r.m.s. $\deltat$ resolution is 1.1ps for
$\babar$ and 1.43ps for Belle, both obtained from data.

\subsection{Flavor Tagging}

The flavor of the $\Btag$ is determined through various flavor signatures among 
its daughter tracks. High momentum (primary) leptons, kaons and 
soft pions from $\Dstarp$ decay are primary sources for flavor tagging. 
In addition, $\Lambda$ baryons and lower momentum (secondary) leptons 
can also be used to assist tagging. To obtain optimal tagging efficiency, both experiments 
use multivariate algorithms to combine various sources of flavor information in an event.  
Similar events, judged by their physics content or estimated tagging purity, are usually grouped into 
tagging categories to aid in the study of tagging-based systematic errors.

The figure of merit for B flavor-tagging is the effective tagging efficiency, 
\begin{eqnarray}
Q \equiv \sum_i {\eps_i (1-2\mistag_i)^2},
\end{eqnarray}
where $i$ sums over tagging categories. Since the \CP measurement error
and tagging efficiency are related ($\sigma_{asym} \propto 1/\sqrt{Q}$),
a higher effective tagging efficiency reduces \CP measurement error.

At $\babar$, events are grouped into four hierarchical, mutually exclusive  
tagging categories based on their physics contents. The $\leptontag$  category
contains events with an identified high momentum lepton. Events with a kaon are assigned to 
either the $\kaonitag$ or $\kaoniitag$  category. Among the two, the $\kaonitag$ category 
contains events with higher estimated tagging probability, contributed by additional tagging 
sources such as  a soft pion compatible  with $\Dstarp$ decay. The $\kaoniitag$ category also contains remaining 
events with a soft pion. All other events are assigned to the $\othertag$
category except for those that have no useful tagging information (which are  
excluded from further analysis). 
A set of neural networks have been developed to classify events and to provide
estimated mistag probability. The efficiency and mistag probability for each 
of the four tagging categories can be obtained from data as shown in table~\ref{tab_mistag_BaBar}.  Based on these measured efficiencies and mistag probability, the effective tagging efficiency($Q$) is calculated to be $28.1\%$. 

\begin{table}[!t]
\centering
\caption
{Efficiencies $\epsilon_i$, average mistag fractions $\mistag_i$, mistag fraction differences
$\Delta\mistag_i=\mistag_i(\Bz)-\mistag_i(\Bzb)$, and $Q$ extracted for each tagging
category $i$ from the $B_{\rm flav}$ and $B_{\CP}$ sample. This data was collected
by the $\babar$ collaboration.
}
\label{tab_mistag_BaBar} 
\begin{tabular}{|c|c|c|c|l|}
\hline
Category     & $\varepsilon$   (\%)& 
               $\mistag$       (\%)& 
               $\Delta\mistag$ (\%)&
               $Q$             (\%) \\   
\hline
\leptontag   &$9.1\pm0.2$ & $3.3\pm 0.6$ & $-1.5\pm1.1 $&$7.9\pm0.3$  \\  
\hline
\kaonitag    &$16.7\pm0.2$ & $10.0\pm 0.7$ & $ -1.3\pm1.1 $&$10.7\pm0.4$  \\ 
\hline
\kaoniitag   &$19.8\pm0.3$ & $20.9\pm 0.8$ & $ -4.4\pm1.2 $&$6.7\pm0.4$  \\ 
\hline
\othertag    &$20.0\pm0.3$ & $31.5\pm 0.9$ & $ -2.4\pm1.3 $&$2.7\pm0.3$  \\   
\hline
All          &$65.6\pm0.5$ &           &          &$28.1\pm0.7$  \\ 
\hline
\end{tabular} 
\end{table} 

At Belle, events are instead grouped into tagging categories based solely on estimated 
tagging probability. A quantity $r$ is assigned to each event. 
An $r$ value of zero signifies no tagging power and an $r$ value of 1 means perfect tagging.
Events are sorted into six intervals of $r$ between 0 and 1, according 
to flavor purity. The event fraction and mistag probability for each category are 
determined directly from data as summarized in table~\ref{tab_mistag_Belle}. 
The corresponding $Q$ value for Belle is $28.8\%$, similar to that of $\babar$'s.        

\begin{table}[hbtp]
\centering
\caption{The event fractions ($\epsilon_l$) and wrong tag
fractions ($w_l$) for each $r$ interval. The errors include
both statistical and systematic uncertainties. The
event fractions are obtained from the $\jpsi \KS$ simulation. This data
was collected by the Belle collaboration.}
\begin{tabular}{|c|c|c|l|}
\hline
$Category(l)$ & $r$ & $\epsilon_l$ &$w_l$ \\
\hline
1 & 0.000 $-$ 0.250 & 0.399 & $0.458\pm0.006$ \\
\hline
2 & 0.250 $-$ 0.500 & 0.146 & $0.336\pm0.009$ \\
\hline
3 & 0.500 $-$ 0.625 & 0.104 & $0.229~^{+0.010}_{-0.011}$ \\
\hline
4 & 0.625 $-$ 0.750 & 0.122 & $0.159\pm0.009$ \\
\hline
5 & 0.750 $-$ 0.875 & 0.094 & $0.111\pm0.009$ \\
\hline
6 & 0.875 $-$ 1.000 & 0.137 & $0.020~^{+0.007}_{-0.006}$ \\
\hline
\end{tabular}
\label{tab_mistag_Belle}
\end{table}

%%%%%%%%%%%%%%%%%%%%%%%%%%%%%%%%%%%%%%%%%%%%%%%%%%%%%%%%%%%%%%%%%%%%%%%%
%%%%%%%%%%%%%%%%%%%%%%%%%%%%%%%%%%%%%%%%%%%%%%%%%%%%%%%%%%%%%%%%%%%%%%%%
%%%%%%%%%%%%%%%%%%%%%%%%%%%%%%%%%%%%%%%%%%%%%%%%%%%%%%%%%%%%%%%%%%%%%%%%

\section{Measurement Of $\BzBzb$ Oscillation Frequency}
To measure the $\Bz$ mixing parameter $\deltamd$, the flavors of both the $\Brec$ and 
$\Btag$ need to be determined. The mixing frequency is extracted from the time evolution of 
opposite-flavor (``unmixed'') and same-flavor (``mixed'') B-decays. 
The physics probability density function (PDF), before accounting for 
detector and background effects, is:
\begin{eqnarray}
{\rm f}(\deltat) &=& {\frac{e^{{- \left| \deltat \right|}/\tau_{\Bz} }}{4\tau_{\Bz} }}[ 1 \pm  \cos{( \deltamd  \deltat )}],
\label{eq:mixingpdf}
\end{eqnarray}
where $\tau_{\Bz}$ is the \Bz lifetime, and ``$\pm$'' denotes ``$+$'' for unmixed events and ``$-$'' for mixed  events.

Samples that can be used for the mixing measurement include: 
\begin{enumerate}
\setlength{\parskip}{0pt}\setlength{\itemsep}{0pt}
\item ``Fully Hadronic'', where $\Brec$ is completely reconstructed to the exclusive
 hadronic decays  $\ddstarh$ (where $h=\pip,\rhop,\aonep$), 
$\jpsi\Kstarz (\Kstarz\to \Kp\pim)$. 

\item ``Semileptonic'', where $\Brec$ is reconstructed to $\Dstarm \lp \nu$. 

\item ``Partial $\Dstar\pi$'', where $\Brec$ is partially reconstructed to $\Dstarm\pip$ ($\Dstarm\to\Dz\pim$).

\item ``Dilepton'', where instead of attempting to reconstruct one of the B decays, events with two high momentum leptons are used.

\end{enumerate}

For the first three samples above, the flavor of the $\Brec$ is determined by the 
charge of its daughters and the flavor of the $\Btag$ is provided by flavor tagging. 
The time-difference $\deltat$ is determined using the $\Brec$ and $\Btag$ vertices. 
For dilepton samples, the charges of the two leptons (which are presumed to be from semileptonic
B decays) indicate the flavor of the B-mesons. Proper-time information is obtained 
using the impact parameters of the two leptons.

\noindent
BaBar has reported results from three measurements~\cite{bib:mixing_babar}:
\begin{itemize}
\setlength{\parskip}{0pt}\setlength{\itemsep}{0pt}
\item $\deltamd = 0.516 \pm 0.016 \stat \pm 0.010 \syst \invps$ (``Hadronic'' , 30 $\invfb$)
  
\item $\deltamd = 0.492 \pm 0.018 \stat \pm 0.013 \syst \invps$ (``Semileptonic'', 21 $\invfb$)
  
\item $\deltamd = 0.493 \pm 0.012 \stat \pm 0.009 \syst \invps$ (``Dilepton'', 21 $\invfb$)
\end{itemize}  
%
%These measurements give a combined $\babar$ result of
%
%\begin{eqnarray*}
%\deltamd = 0.500 \pm 0.008 \stat \pm 0.006 \syst \invps
%\end{eqnarray*}
% 
Belle has reported results from four measurements~\cite{bib:mixing_belle}:

\begin{itemize}
\setlength{\parskip}{0pt}\setlength{\itemsep}{0pt}
\item $\deltamd = 0.528 \pm 0.017 \stat \pm 0.011 \syst \invps$ (``Hadronic'', 29 $\invfb$)
  
\item $\deltamd = 0.494 \pm 0.012 \stat \pm 0.015 \syst \invps$ (``Semileptonic'', 29 $\invfb$)
  
\item $\deltamd = 0.505 \pm 0.017 \stat \pm 0.020 \syst \invps$ (``Partial $\Dstar\pi$''  , 29 $\invfb$)

\item $\deltamd = 0.503 \pm 0.008 \stat \pm 0.009 \syst \invps$ (``Dilepton'', 29 $\invfb$)

\end{itemize}  
%
%These measurements give a combined Belle result of
%
%\begin{eqnarray*}
%\deltamd = 0.506 \pm 0.006 \stat \pm 0.007 \syst \invps
%\end{eqnarray*}
% 
Combining $\babar$ and Belle results yields $\deltamd = 0.503\pm0.007 \invps$, 
as compared with combined non B-factory results $\deltamd = 0.498\pm0.013 \invps$ (LEP+SLD+CDF). 

If all results are combined, a world average $\deltamd$ value of  $0.503 \pm 0.006 \invps$ is obtained.

%%%%%%%%%%%%%%%%%%%%%%%%%%%%%%%%%%%%%%%%%%%%%%%%%%%%%%%%%%%%%%%%%%%%%%%%
%%%%%%%%%%%%%%%%%%%%%%%%%%%%%%%%%%%%%%%%%%%%%%%%%%%%%%%%%%%%%%%%%%%%%%%%
%%%%%%%%%%%%%%%%%%%%%%%%%%%%%%%%%%%%%%%%%%%%%%%%%%%%%%%%%%%%%%%%%%%%%%%%

\section{CP Violation Measurement With Charmonium $\b \to\ccbar s$ Final States}
For the measurement of CP asymmetries, the $\Brec$ needs to be 
reconstructed in a CP eigenstate ($B_{CP}$) with eigenvalue $\eta_f=-1$ or $+1$.  
For events where $\Brec$= $B_{CP}$ and the flavor of $\Btag$ is known to be $\Bz (\Bzb)$, 
the decay rate 
 ${\rm f}_+({\rm f}_-)$ is given by 
\begin{equation}
{\rm f}_\pm(\deltat) = {\frac{e^{{- \left| \deltat \right|}/\tau_{\Bz} }}{4\tau_{\Bz} }}  
\Bigg[ 1 \pm {\frac{2\mathop{\cal I\mkern -2.0mu\mit m}\lambda}{1 + |\lambda|^2} }\sin{( \Delta m_{d}  \deltat )} \mp {\frac{1  - |\lambda|^2 } {1 + |\lambda|^2} } \cos{( \Delta m_{d}  \deltat) }  \Bigg],
\label{eq:timedist}
\end{equation}
where $\lambda$ is  a complex parameter 
that depends on both the \Bz-\Bzb oscillation amplitude and the amplitudes
describing \Bzb and \Bz decays to a common \CP final state.
 $\CP$ violation arises  if $\lambda$ is not unity. In other words, $\CP$ violation is manifested with a non vanishing sine or cosine term in the equation. Experimentally, \CP violation can be observed as a difference between the \deltat distributions of \Bz- and \Bzb-tagged events or as an asymmetry with respect to $\deltat = 0$ for either flavor tag. 

\par

Among many possible CP modes,  $b\to\ccbar s$ (charmonium) decays offer the 
best opportunity for CP violation measurement~\cite{bib:BCP}. These modes include the CP-odd ($\eta_f=-1$)
final states $\jpsi\KS$, $\psitwos\KS$, $\chicone \KS$, and $\eta_c \KS$, and 
CP-even ($\eta_f=+1$) state $\jpsi\KL$. In addition, a CP-mixed state $\jpsi\Kstarz$,
where $\Kstarz$ decays to $\KS\piz$, can also be used after its CP 
composition is measured through an angular analysis. For this CP-mixed 
$\jpsi\Kstarz$ decay, $\babar$ and Belle find the CP-odd fraction to be $16.0\pm 3.5\%$
and $19\pm 2\stat\pm 3\syst\%$, respectively. This fraction can be used to 
compute an effective $\eta_f$ ($\sim 0.65$) for use in the CP extraction.

%\begin{table}[b] 
\begin{table}[htb]
\footnotesize
\caption{ 
Number of signal events $N_{\rm tag}$ after tagging and 
vertexing requirements, signal purity $P$, and results of fitting for \CP\ asymmetries in the $B_{\CP}$ sample and in various sub-samples, 
as well as in the $B_{\rm flav}$ and charged $B$ control samples.  
Errors are statistical only. ($\babar$ experiment)}
\label{tab:yields_BaBar} 
\begin{center}
\begin{tabular}{|l|c|c|c|} \hline
 Sample  & $N_{\rm tag}$ & $P(\%)$ & $\stwob$ \\ \hline
$\jpsi\KS$,$\psitwos\KS$,$\chicone\KS$,$\etac\KS$   & $1506$        & $94$       &  $0.76\pm 0.07 $  \\
$\jpsi \KL$ $(\eta_f=+1)$                           & $988$        & $55$       &  $0.72\pm 0.16 $  \\
$\jpsi\Kstarz (\Kstarz \to \KS\piz)$                 & $147$         & $81$       &  $0.22\pm 0.52$   \\
\hline
$\ \ \jpsi \KS$ ($\KS \to \pi^+ \pi^-$)    & $974$        & $97$       &  $0.82\pm 0.08$ \\
$\ \ \jpsi \KS$ ($\KS \to \pi^0 \pi^0$)    & $170$        & $89$       &  $0.39\pm 0.24$ \\
$\ \ \psi(2S) \KS$ ($\KS \to \pi^+ \pi^-$) & $150$        & $97$       &  $0.69\pm 0.24$ \\
$\ \ \chicone \KS $                        & $80$         & $95$       &  $1.01\pm 0.40$ \\
$\ \ \etac\KS $                            & $132$        & $73$       &  $0.59\pm 0.32$ \\
\hline
$\ $ \leptontag\ category                & $220$        & $98$       &  $0.79\pm 0.11$   \\
$\ $ \kaonitag\ category                 & $400$        & $93$       &  $0.78\pm 0.12$   \\
$\ $ \kaoniitag\ category                & $444$        & $93$       &  $0.73\pm 0.17$   \\
$\ $ \othertag\ category                 & $442$        & $92$       &  $0.45\pm 0.28$   \\
\hline
$\ $ \Bz\ tags                           & $740$        & $94$       &  $0.76\pm 0.10 $  \\
$\ $ \Bzb\ tags                          & $766$        & $93$       &  $0.75\pm 0.10 $  \\
\hline\hline
$B_{\rm flav}$ sample                    & $25375$      & $85$       &  $0.02\pm 0.02$   \\
\hline 
charged $B$ sample                             & $22160$      & $89$       &  $0.02\pm 0.02$   \\ 
\hline
\hline
 Full \CP\ sample                        & $2641$        & $78$       &  $0.74 \pm 0.07 $  \\
\hline 
\end{tabular} 
\end{center}
\end{table}

\begin{table}[thb]
\footnotesize
\caption{The numbers of reconstructed $B \to f_{CP}$
candidates before flavor tagging and vertex reconstruction
($N_{\rm rec}$),
the numbers of events used for the $\sin 2\phi_1$ determination
($N_{\rm ev}$), and the
estimated signal purity for each $f_{CP}$ mode. (Belle experiment.)}
\begin{center}
\begin{tabular}{|l|cc|c|c|}
\hline
Sample & $N_{\rm rec}$ & $N_{\rm ev}$ & Purity &$\stwofone$ \\
\hline
$\jpsi(\ell^+\ell^-)\KS(\pi^+\pi^-)$ &  1285 & 1116 & 0.98 &$0.73\pm 0.10$\\
\hline
$\jpsi(\ell^+\ell^-)\KS(\pi^0\pi^0)$ & 188 & 162   & 0.82 & \\
\cline{1-4}
$\psi(2S)(\ell^+\ell^-)\KS(\pi^+\pi^-)$  & 91 & 76    & 0.96 &\\
\cline{1-4}
$\psi(2S)(\jpsi\pi^+\pi^-)\KS(\pi^+\pi^-)$ & 112 & 96    & 0.91 & \\
\cline{1-4}
$\chi_{c1}(\jpsi\gamma)\KS(\pi^+\pi^-)$ &  77 & 67    & 0.96 & $0.67\pm 0.17$\\
\cline{1-4}
$\eta_c(\KS K^-\pi^+)\KS(\pi^+\pi^-)$ &  72 & 63    & 0.65 &\\
\cline{1-4}
$\eta_c(K^+K^-\pi^0)\KS(\pi^+\pi^-)$ &  49 & 44    & 0.72 &\\
\cline{1-4}
$\eta_c(p\overline{p})\KS(\pi^+\pi^-)$      &  21 & 15    & 0.94 &\\
\hline
$\jpsi(\ell^+\ell^-) K^{*0}(\KS\pi^0)$& 101 & 89    & 0.92 &$0.04\pm 0.63$\\
\hline
$\jpsi(\ell^+\ell^-) \KL$                & 1330 &1230  & 0.63 &$0.78\pm 0.17$\\
\hline \hline
All CP Sample                        & 3326 &2958  & 0.81 &$0.72\pm 0.07$\\
\hline
\end{tabular}
\end{center}

\label{tab:yields_Belle}
\end{table}

In the Standard Model, $\lambda$ is expected to be $\eta_f e^{-2i\beta}$ ($|\lambda|=1$, $Im(\lambda)=-\eta_f\stwob$) for
these charmonium decays.  Thus, a measurement with the time dependent decay 
rates in equation~\ref{eq:timedist} directly reveals the CP parameter 
$\stwob$ with little ambiguity. 

Both $\babar$ and Belle reconstruct these charmonium modes for use in their 
$\stwob$ measurements~\cite{bib:ichep02_stwob_babar,bib:ichep02_stwob_belle}. 
Yields on each signal mode are summarized in 
table~\ref{tab:yields_BaBar} for BaBar~\footnote{Only events with a flavor tag are 
included, total Tagging efficiency is $65.6\pm 0.5\%$.}  and in table~\ref{tab:yields_Belle} for Belle.
Shown also in these two tables are measured $\stwob$ ($\stwofone$) value for each sub-sample (see text below).

After the flavor of the $\Btag$ and time-difference $\deltat$  are 
determined for each event in the CP sample, the whole sample is 
used to construct a likelihood function based on the PDF 
\begin{equation}
{\rm f}(\deltat) = {\frac{e^{{- \left| \deltat \right|}/\tau_{\Bz} }}{4\tau_{\Bz} }}[ 1 - \eta_f q(1-2w) \stwob \sin{( \deltamd  \deltat )}]
%{\rm f}(\deltat) = {\frac{e^{{- \left| \deltat \right|}/\tau_{\Bz} }}{4\tau_{\Bz} }}  \Bigg[{ 1 - \eta_f q(1-2w) \stwob \sin{( \deltamd  \deltat )}}  \Bigg],
\label{eq:sin2bpdf}
\end{equation}
where $q = +1(-1)$ when $\Btag$ is tagged as $\Bz$ ($\Bzb$) and $w$ 
is the estimated mistag probability for the tagging category to which the event belongs. 
As mentioned earlier, both $\babar$ and Belle obtain $w$ from data. 
The CP-parameter $\stwob$ in the PDF serves as a free parameter and is to be 
extracted from a fit on the data employing the PDF.

The above physics PDF has to be modified to take into account the time 
resolution function and background time distribution. Details of time
resolution treatment and fitting procedure can be found 
in ~\cite{bib:babar-stwob-prd} for $\babar$ 
and in ~\cite{bib:belle-stwob-prd} for Belle.

The value of $\stwob$ is determined by an unbinned 
maximum-likelihood fit to the observed $\deltat$ distribution. 
For all CP modes combined, the fitted $\stwob$ ($\stwofone$) values are:
\begin{eqnarray*}
\stwob &=& 0.741 \pm 0.067\ \stat \pm 0.033\ \syst. \, \mathrm{(BaBar)} \\
\stwofone &=& 0.719 \pm 0.074\ \stat \pm 0.035\ \syst. \, \mathrm{(Belle)}
\end{eqnarray*}

Fitted $\stwob$ values for various sub-samples are  included in 
table~\ref{tab:yields_BaBar} for $\babar$ and table~\ref{tab:yields_Belle} 
for Belle. No inconsistency between the samples is observed.

Combining these latest two $\stwob$ results from $\babar$ and Belle with earlier 
(non B-factory) results, namely ($0.84^{+0.82}_{-1.04}\pm 0.16$) from Aleph, 
($0.79^{+0.41}_{-0.44}$) from CDF and ($3.20^{+1.8}_{-2.0}\pm 0.5$) from OPAL, 
a world average of $\stwob=0.735 \pm 0.055 $ is obtained. 

\begin{figure}[htb]
\begin{center}
\includegraphics[width=0.6\textwidth,clip]{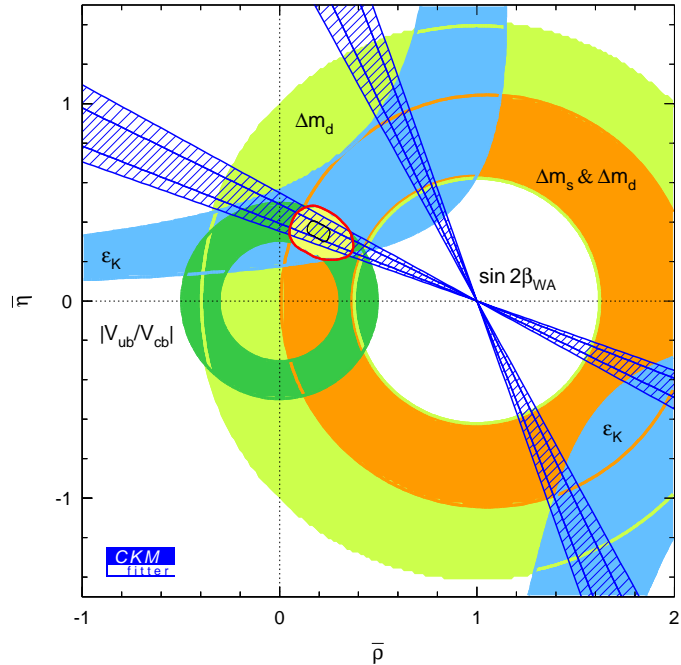}
\end{center}
\caption{Directly measured  $\stwob$ result shown as straight lines in the 
CKM \etabar-\rhobar plane.   Contours near $\etabar=0.3\ and\ \rhobar=0.2$ 
are Standard Model prediction fitted with constraints from other measurements.}
\label{fig:ckm_sin2b_wa}
\end{figure}

This world average value on (directly measured) $\stwob$ can be compared with
the Standard Model constraints in the $\rhobar-\etabar$ plane, as shown in
figure~\ref{fig:ckm_sin2b_wa}. The indirect constraints are realized from measurements on 
$\vubvcb$, $\deltamd$, $\deltams$ and CP-violation in the kaon system. Within 
the current measurement uncertainties, good agreement is observed. Measurements 
with improved precision and, in particular, measurements of other CP angles 
are necessary to provide a more stringent test on the Standard Model CKM 
theory of CP-violation.      

The $\stwob$ measurements discussed so far are performed with the assumption 
$|\lambda|=1$, as is predicted by the Standard Model for the  $b\to\ccbar s$ decays. 
To test this assumption, a more general physics PDF, shown earlier in equation~\ref{eq:timedist},
can instead be used. A fitting based on this generalized PDF gives~\footnote{Belle uses all \CP modes in this generalized fit while $\babar$ fits only on \CP-odd modes.} 
\begin{eqnarray*}
\vert\lambda\vert &=& 0.948 \pm 0.051\ \stat \pm 0.017\ \syst \, \mathrm{(BaBar)}\\
\vert\lambda\vert &=& 0.950 \pm 0.049\ \stat \pm 0.026\ \syst \, \mathrm{(Belle)}
\end{eqnarray*} 
The coefficient of $\sin(\deltamd\deltat)$ term is simultaneously fitted to be 
$0.759 \pm 0.074\ \stat$ at $\babar$ and $0.720 \pm 0.074\ \stat$ at Belle, 
respectively. These results are consistent with the original assumption of $|\lambda|=1$.  

\section{CP Violation Measurement With Other Modes}
In addition to the $b\to\ccbar s$ charmonium modes, \CP violation measurement 
can be performed with many other \CP decays. In this section, \CP results measured from two classes of $\stwob$~\footnote{Results on $\stwoa$ are summarized in a separate article in this proceedings.} sensitive samples, the Cabibbo-suppressed 
$b\to\ccbar d$ decays and the penguin dominated $b\to\ssbar s$ decays, are briefly summarized. More details can be found in ~\cite{bib:ichep02_othercp_babar, bib:ichep02_othercp_belle}.

Unlike the theoretically clean $b\to\ccbar s$ decays with which $\stwob$ can be directly measured, these additional modes may be affected by more than 
one \CP-violating phases. The Standard Model assumption 
$\lambda=\eta_f e^{-2i\beta}$ can not always be applied; often, a generic 
form of physics PDF as defined in equation~\ref{eq:timedist} 
has to be used.  With this generic PDF, \CP asymmetry coefficients $S_f (\equiv \frac{2\mathop{\cal I\mkern -2.0mu\mit m}\lambda}{1 + |\lambda|^2})$   and  
$C_f (\equiv \frac{1  - |\lambda|^2 } {1 + |\lambda|^2})$ can be extracted and compared with theoretical predictions. In the limit that only one weak phase 
contributes, the coefficient $S_f$ should be equal to $-\eta_f\stwob$, 
where $\eta_f$ is the eigen value of the corresponding \CP mode, and the coefficient $C_f$ should be equal to zero. 
            
\subsection{Time Dependent CP Asymmetries With $b \to\ccbar d$ Decays}
One useful $b\to\ccbar d$ mode is the  $\Bz \to \jpsi\piz$ decay where a \CP even ($\eta_f=+1$) final state is produced. The decay process receives both tree and penguin contributions. The Cabibbo-suppressed tree contribution has the same weak phase as  the $b \to\ccbar s$ modes but the penguin contribution of comparable strength may bring in a different weak phase. Both $\babar$ and Belle have reconstructed events in this mode; the measured \CP asymmetry coefficients are:
\begin{eqnarray*}
S_{\jpsi\piz} &=& 0.05 \pm 0.49\ \stat \pm 0.16\ \syst \, (\babar) \\
 C_{\jpsi\piz} &=& 0.38 \pm 0.41\ \stat \pm 0.09\ \syst \, (\babar)  \\ 
S_{\jpsi\piz} &=&\  0.93 \pm 0.49\ \stat \pm 0.08\ \syst \, \mathrm{(Belle)} \\C_{\jpsi\piz} &=&-0.25 \pm 0.39\ \stat \pm 0.06\ \syst \, \mathrm{(Belle)}.
\end{eqnarray*} 

In addition to $\Bz \to \jpsi\piz$, $\babar$ has  also constructed the 
$\Bz \to \Dstarp\Dstarm$ decay. This decay is also a $\b \to\ccbar d$ 
process but the final state $\Dstarp\Dstarm$ is not a \CP eigenstate -- an angular analysis is necessary  to determine the \CP composition. With their $\Dstarp\Dstarm$ sample, $\babar$ has measured a \CP-odd  fraction of $0.096 \pm 0.060 \stat$  and extracted the effective \CP asymmetry parameters as:
\begin{eqnarray*}
Im(\lambda) &=& 0.31 \pm 0.43\ \stat \pm 0.13\ \syst  \\
|\lambda| &=& 0.98 \pm 0.25\ \stat \pm 0.09\ \syst    
\end{eqnarray*} 
If the $\Bz \to \Dstarp\Dstarm$ decay is a tree-only process,  $Im(\lambda)=-\stwob$ and $|\lambda|=1$ are expected.  In the Standard Model, penguin-induced correction is predicted to be small ($< 2\%$) compared to this tree-only \CP assymmetry.

\subsection{Time Dependent CP Asymmetries With $b\to\ssbar s$ Decays}
The penguin dominated $b\to\ssbar s$ process is also sensitive to $\stwob$. If only the Standard Model weak phase contributes, the \CP coefficients $S_f$ and $C_f$ are expected to be $-\eta_f\stwob(\stwofone)$ and zero, respectively. Significant deviations to these expected value probe for new physics (in the penguin loops, for example). 

$\babar$ has reported \CP results from the \CP-odd $\Bz \to \phi \KS$ decays. 
The  effective $\stwob$ value is found to be $-0.19^{+0.52}_{-0.50} \stat \pm 0.09 \syst$.

Belle has presented results with two \CP-odd, $\Bz \to \phi \KS$ and $\Bz to \etapr \KS $ modes, and a \CP mixed mode\footnote{For this \CP mixed mode, Belle has measured its \CP composition to be 97\% \CP-odd and 3\% \CP-even.}  
$\Bz \to \Kp \Km \KS$ as well.
The ``$\stwofone$'' $(\equiv -\eta_fS_f)$ valuses measured from these three decays are $0.76 \pm 
0.36 \stat ^{+0.05}_{-0.06} \syst$, $-0.73 \pm 0.64  \stat \pm 0.18 \syst$, 
and $0.52 \pm 0.46 \stat \pm 0.11 \syst$, respectively~\footnote{The systematic error for the $\Kp \Km \KS$ mode is subject to an additional $^{+0.27}_{-0.03}$ contribution from the uncertainty in the fraction of the \CP-odd component}. In the meantime, \CP asymmetry parameter $C_f$ is also measured with each mode
and found to be consistent with zero.

\section{Acknowledgments}
It is a pleasure to thank the organizers of the XXII Physics In Collision Conference for their kind invitation to present this review. This work is not possible without the aid of many individuals in the $\babar$ and the Belle Collaborations who provided me their latest results. I am grateful to   P. Burchat, M. Hazumi, A. Jawahery, Y. Sakai, S. Sekula, S. L. Wu, and many others for 
their very helpful inputs and comments. This work is supported by the US Department of Energy contract DE-FG0295-ER40896.


\begin{thebibliography}{99}

\bibitem{bib:CKM}
\hyphenation{Ko-ba-ya-shi}
N.~Cabibbo, \prl {\bf 10}, 531 (1963); M.~Kobayashi and T.~Maskawa, \progtp {\bf 49}, 652 (1973).

\bibitem{bib:babar-stwob-2001}
B.\ Aubert {\em et al.} (\babar\ Collab.), ,
\prl {\bf 87}, 091801 (2001).

\bibitem{bib:belle-stwob-2001}
K.\ Abe {\em et al.} (Belle Collab.),
\prl {\bf 87}, 091802 (2001).


\bibitem{bib:babar} 
 B.\ Aubert {\em et al.} ($\babar$ Collab.), \nim{\bf A479}, 1 (2002).

\bibitem{bib:belle}
A.~Abashian {\it et al.} (Belle Collab.),
Nucl. Instr. and Meth. A {\bf 479}, 117 (2002).

\bibitem{bib:kekb}
 E.~Kikutani ed., KEK Preprint 2001-157 (2001),
 to appear in Nucl. Instr. and Meth. A.


\bibitem{bib:mixing_babar}
B.\ Aubert {\em et al.} ($\babar$ Collab.), \prl {\bf 88}, 221802 (2002);
B.\ Aubert {\em et al.} ($\babar$ Collab.), \prl {\bf 88}, 221803 (2002);
B.\ Aubert {\em et al.} ($\babar$ Collab.), hep-ex/02070071.

\bibitem{bib:mixing_belle}
K.~Abe {\it et al.} (Belle Collab.), hep-ex/0207022;
K.~Abe {\it et al.} (Belle Collab.), BELLE-CONF 0203, hep-ex/0207045;
K.~Abe {\it et al.} (Belle Collab.), BELLE-CONF 0204;
K.~Abe {\it et al.} (Belle Collab.), BELLE-CONF 0205.

\bibitem{bib:BCP}
A.B.~Carter and A.I.~Sanda, \pr {\bf D23}, 1567 (1981);
%\\
I.I.~Bigi   and A.I.~Sanda, \np {\bf B193}, 85 (1981).

\bibitem{bib:babar-stwob-prd}
\babar\ Collaboration, B.\ Aubert {\em et al.} ($\babar$ Collab.), 
SLAC-PUB-9060, hep-ex/0201020, to appear in \prd .

\bibitem{bib:belle-stwob-prd}
K.~Abe {\it et al.} (Belle Collab.),  Phys. Rev. Lett. {\bf 87},
091802 (2001); K.~Abe {\it et al.} (Belle Collab.), hep-ex/0202027, accepted for publication in Phys. Rev. D.

\bibitem{bib:ichep02_stwob_babar}
B.\ Aubert {\em et al.} ($\babar$ Collab.), SLAC-PUB-9293, hep-ex/0207042;

\bibitem{bib:ichep02_stwob_belle}
K.~Abe {\it et al.} (Belle Collab.), hep-ex/0207098;

\bibitem{bib:ichep02_othercp_babar}
B.\ Aubert {\em et al.} ($\babar$ Collab.), SLAC-PUB-9297, hep-ex/0207070;
B.\ Aubert {\em et al.} ($\babar$ Collab.), SLAC-PUB-9298, hep-ex/0207058;
B.\ Aubert {\em et al.} ($\babar$ Collab.), SLAC-PUB-9299, hep-ex/0207072;

\bibitem{bib:ichep02_othercp_belle}
K.-F.~Chen {\it et al}, hep-ex/0207033, to appear in 
{\em Phys.~Lett.~B} (2002);
K.~Abe {\it et al.}, BELLE-CONF-0232 (2002);
K.~Abe {\it et al.}, BELLE-CONF-0209 (2002);
K.~Abe {\it et al.}, BELLE-CONF-0225 (2002).

\end{thebibliography}
\end{document}